\documentclass[12pt]{article}
\usepackage[paper=a4paper,margin=1in]{geometry}
\usepackage{t1enc}      

\usepackage{amsmath}
\usepackage{amsfonts}
\usepackage{amssymb}
\usepackage{amsthm}
\usepackage{graphicx}
\usepackage{mathrsfs}  

\newcommand{\bi}{\bf i}
\newcommand{\bj}{\bf j}
\newcommand{\bk}{\bf k}

\newtheorem{proposition}{Proposition}

\newtheorem*{theorem*}{Theorem}

\providecommand{\keywords}[1]
{\small	\textbf{\textit{Keywords:}} #1 }

\numberwithin{equation}{section}

\begin{document}
\bibliographystyle{unsrt}

\title{A note on Penrose's Spin-Geometry Theorem and the 
geometry of `empirical quantum angles'}
\author{L\'aszl\'o B. Szabados \\
Wigner Research Centre for Physics, \\
H-1525 Budapest 114, P. O. Box 49, EU \\ 
E-mail: lbszab@rmki.kfki.hu}

\maketitle

\begin{abstract}
In the traditional formalism of quantum mechanics, a simple direct proof of 
(a version of) the Spin Geometry Theorem of Penrose is given; and the 
structure of a model of the `space of the quantum directions', defined in 
terms of elementary $SU(2)$-invariant observables of the quantum mechanical 
systems, is sketched. 
\end{abstract}

\keywords{spin networks, Spin-Geometry Theorem, quantum geometry, empirical 
angle}


\section{Introduction}
\label{sec-1}

Penrose, in the pioneering work \cite{Pe79} (which appeared first already 
in 1966 as the appendix of his Adams Prize essay), suggested the so-called 
$SU(2)$ spin network as a simple model for the quantum spacetime (see also 
\cite{Pe71a,Pe10,Pe71b,Pe72}). The key idea was that, unifying and splitting 
quantum mechanical systems and measuring the probabilities of occurrence 
of the various total (i.e. the $j$) values of angular momenta obtained in 
this procedure, an \emph{empirical angle} between the angular momentum 
vectors can be \emph{defined}; and the key result was that, in the large 
$j$ (i.e. in the classical) limit, these angles tend to be angles between 
directions in the Euclidean 3-space. 

The significance of this result is that the (conformal structure of the) 
`physical 3-space' that we use as an \emph{a priori} given `arena' in which 
the physical objects are thought to be arranged and the interactions between 
them occur is \emph{determined by the quantum physical systems themselves 
in the classical limit}. Its proof was based on combinatorial/graphical 
techniques. The key idea and a sketch of the proof appeared in 
\cite{Pe10,Mo}, but so far the detailed and complete proof has not been 
published \cite{Pe22}: the proof in \cite{Pe10} remained incomplete. Later, 
another (and \emph{mathematically} different) version of the \emph{same 
physical result} became known as the Spin Geometry Theorem \cite{Mo}. The 
proof of the latter version was based on the more familiar formalism of 
quantum mechanics. (For some more historical remarks, see also \cite{Pe10}.) 

However, soon after the appearance of the idea how the `true' geometry of 
space(time) should be \emph{defined in an operational way}, the emphasis was 
shifted from the systematic investigation of the \emph{consequences} of the 
original ideas to the development of \emph{mathematical theories} of 
possible \emph{a priori quantum geometries}. The latter were only motivated 
by, but were not based directly on the original ideas. The twistor theory 
\cite{PeMacC,BaBa90,Atiyahetal}, and also canonical quantum theories of 
gravity (see e.g. \cite{RoSm90}) are such promising mathematical models. In 
particular, in the latter, the area and volume \cite{RoSm95a} and also the 
angle \cite{Ma99b}, represented by appropriately regularized quantum 
operators, have been shown to be discrete; and, remarkably enough, the spin 
networks emerged as a basis of the states in this theory \cite{RoSm95b}. 

In the present note, we return to the original idea formulated in 
\cite{Pe79}, and especially in \cite{Pe71a,Pe10}: while e.g. the electrons 
or the electromagnetic field are \emph{existing objects} (i.e. `things'), 
by their primary definition, the spacetime and its points, the events, are 
\emph{not}. The events are \emph{phenomena}, and the spacetime, the set of 
them, is only a useful \emph{notion} by means of which the laws of Nature 
can be formulated in a convenient, simple way. Thus, we also share the 
positivistic, Machian view (see \cite{Pe71a,Pe10}) that spacetime and its 
geometry should be \emph{defined} in an operational way \emph{by existing 
material systems}. 

Here, we adopt the idea of the algebraic formulation of quantum theory that 
the quantum system is specified completely if its \emph{algebra of (basic) 
observables} (and, if needed, its representation) is fixed. Then the notion 
and all the structures of space/spacetime should be defined \emph{in terms 
of the observables} of the quantum physical subsystems of the Universe. In 
particular, the angles between `directions' associated with quantum 
mechanical subsystems should also be introduced in this way even though the 
`directions' themselves are \emph{not} defined at all in the classical sense. 
Actually, these subsystems are chosen to be `elementary' in the sense that 
the quantum observables are self-adjoint elements of the \emph{enveloping 
algebra of the $su(2)$ Lie algebra} of the angular momentum operators as the 
basic observables. We can use all the structures on this algebra, but, in 
addition to this, \emph{no a priori} notion of space/spacetime, as an `arena' 
of events, is allowed to be used. It is this general strategy (but replacing 
$su(2)$ by the Lie algebra $e(3)$ of the Euclidean group) that we follow in 
\cite{Sz22} in deriving the \emph{metrical} (rather than only the conformal) 
structure of the Euclidean three-space from elementary quantum systems. 

As far as we know, no complete proof of the Spin Geometry Theorem, even the 
version given by Moussouris in \cite{Mo}, has been published. (The sketch of 
Moussouris's proof was summarized in \cite{Ma99b}.) The present note intends 
to make up this shortage, using an improved version of Moussouris's empirical 
angle between the angular momentum vectors of elementary quantum systems. We 
give a new proof of the theorem in the usual (but slightly more algebraic) 
formulation of quantum mechanics. In proving this theorem, no recoupling is 
needed. The present investigation also casts some light on the nature of the 
`space of empirical quantum directions'. 

In the next section we introduce the key notion, the improved version of the 
empirical angle between the angular momentum vectors of quantum systems. 
Based on this notion, in section \ref{sec-3}, we present the new proof and 
discuss the results. In section \ref{sec-4} we sketch some key properties of 
a classical model of the geometry of empirical quantum angles. 


\section{The empirical angle}
\label{sec-2}

Let $S_{\bi}$, ${\bi}=1,2,...,N$, be \emph{classical mechanical} systems 
whose respective states are completely specified by the real 3-vectors $J^a
_{\bi}$, $a=1,2,3$, called their angular momentum vectors. The length of these 
vectors is $\vert J_{\bi}\vert:=\sqrt{\delta_{ab}J^a_{\bi}J^b_{\bi}}$, and the 
\emph{empirical angle} $\theta_{\bi\bk}$ between $J^a_{\bi}$ and $J^a_{\bk}$ is 
defined (with range $[0,\pi]$) by 
\begin{equation}
\cos\theta_{\bi\bk}:=\frac{\delta_{ab}J^a_{\bi}J^b_{\bk}}{\vert J_{\bi}\vert\,
\vert J_{\bk}\vert}. \label{eq:2.1}
\end{equation}
Clearly, both $\vert J_{\bi}\vert$ and $\cos\theta_{\bi\bk}$ are 
$SO(3)$-invariant. If the direction of the angular momentum vector $J^a_{\bk}$ 
can be obtained from that of $J^a_{\bi}$ by an $SO(3)$ rotation of angle 
$\beta_{\bi\bk}$ in the plane spanned by $J^a_{\bk}$ and $J^a_{\bi}$, then, 
clearly, $\theta_{\bi\bk}=\beta_{\bi\bk}$. Thus, by measuring the 
\emph{$SO(3)$-invariant observables} $\cos\theta_{\bi\bk}$ we can recover the 
angle $\beta_{\bi\bk}$ between the angular momentum vectors of the subsystems, 
defined \emph{in the space of the classical observables}. Next we convey 
these ideas into the quantum theory in a systematic way. We will see that, 
for quantum systems, these two concepts of angle split with far reaching 
consequences. 

Let ${\cal S}_{\bi}$, ${\bi}=1,2,...,N$, be \emph{quantum mechanical} systems, 
whose (normalized) vector states $\phi_{\bi}$ (or, in the bra-ket notation, 
$\vert\phi_{\bi}\rangle$) belong, respectively, to the Hilbert spaces 
${\cal H}_{\bi}$. The general (e.g. mixed) states are density operators on 
them: $\rho_{\bi}:{\cal H}_{\bi}\to{\cal H}_{\bi}$. The basic quantum 
observables are the angular momentum vector operators ${\bf J}^a_{\bi}$ 
satisfying the familiar commutation relations $[{\bf J}^a_{\bi},{\bf J}^b
_{\bi}]={\rm i}\hbar\varepsilon^{ab}{}_c{\bf J}^c_{\bi}$, where 
$\varepsilon_{abc}$ is the alternating Levi-Civita symbol, and we lower and 
raise the Latin indices by the Kronecker delta $\delta_{ab}$ and its 
inverse\footnote{Because of the natural action of $SO(3)$ (i.e. of $SU(2)$ 
in its vector representation) on the real 3-space of the basic quantum 
observables spanned by ${\bf J}^1$, ${\bf J}^2$ and ${\bf J}^3$, apart from 
an overall positive factor $\delta_{ab}$ is in fact a \emph{naturally defined 
3-metric}, which is proportional to the Killing--Cartan metric; and
$\varepsilon_{abc}$ is the corresponding \emph{natural volume 3-form on this 
space}.}. 

The action of $SO(3)$ (or rather of $SU(2)$) on this algebra is given by 
${\bf J}^a_{\bi}\mapsto(R^{-1})^a{}_b{\bf J}^b_{\bi}$. If the $SU(2)$ matrix 
$U^A{}_B$ is parameterized by the familiar Euler angles $(\alpha,\beta,
\gamma)$ according to 
\begin{equation}
U^A{}_B=\left(\begin{array}{ccc}
 \exp\bigl(\frac{\rm i}{2}(\alpha+\gamma)\bigr)\cos\frac{\beta}{2} &
 {\rm i}\exp\bigl(-\frac{\rm i}{2}(\alpha-\gamma)\bigr)\sin\frac{\beta}{2}\\
 {\rm i}\exp\bigl(\frac{\rm i}{2}(\alpha-\gamma)\bigr)\sin\frac{\beta}{2} &
 \exp\bigl(-\frac{\rm i}{2}(\alpha+\gamma)\bigr)\cos\frac{\beta}{2} \\
             \end{array}\right), \label{eq:2.2}
\end{equation}
then the corresponding rotation matrix, $R^a{}_b=-\sigma^a_{AA'}U^A{}_B\bar U
^{A'}{}_{B'}\sigma^{BB'}_b$, is 
\begin{equation}
R^a{}_b=\left(\begin{array}{ccc}
 \cos\alpha\cos\gamma-\sin\alpha\cos\beta\sin\gamma &
 -\sin\alpha\cos\gamma-\cos\alpha\cos\beta\sin\gamma &
 \sin\beta\sin\gamma \\
 \cos\alpha\sin\gamma+\sin\alpha\cos\beta\cos\gamma &
 -\sin\alpha\sin\gamma+\cos\alpha\cos\beta\cos\gamma &
 -\sin\beta\cos\gamma \\
 \sin\alpha\sin\beta & \cos\alpha\sin\beta & \cos\beta 
\end{array}\right). \label{eq:2.3}
\end{equation}
Here $\sigma_a^{AA'}$ are the three non-trivial $SL(2,\mathbb{C})$ Pauli 
matrices (including the factor $1/\sqrt{2}$) according to the conventions of 
\cite{PR1,HT}. (The minus sign in the expression of $R^a{}_b$ is a 
consequence of the convention that, in the present note, we lower and raise 
the Latin indices by the \emph{positive definite} metric $\delta_{ab}$ and its 
inverse, respectively, rather than by the negative definite spatial part of 
the Minkowski metric. The spinor name indices $A,B,...$ are lowered and 
raised by the anti-symmetric Levi-Civita symbol $\varepsilon_{AB}$ and its 
inverse.) For later use, note that $(R^{-1}(\alpha,\beta,\gamma))^a{}_b=(R
(\pi-\gamma,\beta,\pi-\alpha))^a{}_b$. 

Considering the systems ${\cal S}_1$,...,${\cal S}_N$ to be a \emph{single} 
system, the space of the vector (or pure) states of the resulting composite 
system will be ${\cal H}:={\cal H}_1\otimes\cdots\otimes{\cal H}_N$. Its 
elements are general linear combinations of the tensor products $\phi_1
\otimes\cdots\otimes\phi_N$ of the pure states of the subsystems, while a 
general state is given by a density operator $\rho:{\cal H}\to{\cal H}$. The 
operators ${\bf J}^a_{\bi}$ define the operators ${\bf I}_1\otimes\cdots
\otimes{\bf J}^a_{\bi}\otimes\cdots\otimes{\bf I}_N$ on ${\cal H}$, denoted 
for the sake of simplicity also by ${\bf J}^a_{\bi}$. Here ${\bf I}_{\bk}$ is 
the identity operator acting on ${\cal H}_{\bk}$. With these notations, the 
operators ${\bf J}^a_{\bi}{\bf J}^b_{\bk}$ are well defined, and ${\bf J}^a
_{\bi}$ and ${\bf J}^b_{\bk}$ are commuting if ${\bi}\not={\bk}$. 

Next, for ${\bi}\leq{\bk}$, let us form the operator ${\bf J}_{\bi}\cdot
{\bf J}_{\bk}:=\delta_{ab}{\bf I}_1\otimes\cdots\otimes{\bf J}^a_{\bi}\otimes
\cdots\otimes{\bf J}^b_{\bk}\otimes\cdots\otimes{\bf I}_N:{\cal H}\to{\cal H}$. 
For ${\bi}={\bk}$ this will be denoted simply by $({\bf J}_{\bi})^2$. 
Motivated by equation (\ref{eq:2.1}), we define 
\emph{the empirical (quantum) angle} between the subsystems ${\cal S}_{\bi}$ 
and ${\cal S}_{\bk}$ in the pure tensor product state $\phi=\phi_1\otimes
\cdots\otimes\phi_N$ by  
\begin{equation}
\cos\theta_{\bi\bk}:=\frac{\langle\phi\vert{\bf J}_{\bi}\cdot{\bf J}_{\bk}
\vert\phi\rangle}{\sqrt{\langle\phi\vert({\bf J}_{\bi})^2\vert\phi\rangle}
\sqrt{\langle\phi\vert({\bf J}_{\bk})^2\vert\phi\rangle}}=
\frac{\langle\phi_{\bi}\vert{\bf J}^a_{\bi}\vert\phi_{\bi}\rangle\,\delta_{ab}
\,\langle\phi_{\bk}\vert{\bf J}^b_{\bk}\vert\phi_{\bk}\rangle}{\sqrt{\langle
\phi_{\bi}\vert({\bf J}_{\bi})^2\vert\phi_{\bi}\rangle}\sqrt{\langle\phi_{\bk}
\vert({\bf J}_{\bk})^2\vert\phi_{\bk}\rangle}}. \label{eq:2.4}
\end{equation}
Since the absolute value of the expression on the right is not greater than 
one, this can, in fact, be considered to be the cosine of some angle $\theta
_{\bi\bk}$; and for the range of this angle it seems natural to choose $[0,
\pi]$. 

\noindent
Remarks: 
\begin{enumerate}
\item $\cos\theta_{\bi\bk}$ depends only on the states of ${\cal S}_{\bi}$ and 
${\cal S}_{\bk}$, and it is independent of the states of the other subsystems. 
Moreover, using the transformation property ${\bf U}^\dagger{\bf J}^a{\bf U}
=R^a{}_b{\bf J}^b$ of the angular momentum vector operator, it is 
straightforward to check that $\cos\theta_{\bi\bk}$ is $SU(2)$-invariant. 

\item
Since for ${\bi}={\bk}$ one has that ${\bf J}_{\bi}\cdot{\bf J}_{\bi}=
\delta_{ab}{\bf J}^a_{\bi}{\bf J}^b_{\bi}$, which is just the Casimir operator 
$({\bf J}_{\bi})^2$ of $su(2)$ on ${\cal S}_{\bi}$, $\cos\theta_{\bi\bi}=1$ 
holds. Thus the empirical angle between any angular momentum vector and 
itself is always zero, as it could be expected. In the rest of this note, 
we assume that ${\bi}\not={\bk}$. 

\item
It is straightforward to define the angle between two subsystems even 
when the state of the composite system is an entangled state, $\phi=\sum
_{i_1,...,i_N}c^{i_1...i_N}\phi_{i_1}\otimes\cdots\otimes\phi_{i_N}$, or when it is 
a general mixed state, represented by a density operator $\rho:{\cal H}\to
{\cal H}$. In the first case, it is still defined by (\ref{eq:2.4}), while 
in the second by 
\begin{equation*}
\frac{{\rm tr}\bigl(\rho\,{\bf J}_{\bi}\cdot{\bf J}_{\bk}\bigr)}{\sqrt{{\rm tr}
  (\rho({\bf J}_{\bi})^2)}{\sqrt{{\rm tr}(\rho({\bf J}_{\bk})^2)}}}.
\end{equation*}
However, in these cases the states of the individual constituent subsystems 
would be \emph{mixed}, and hence the interpretation of $\cos\theta_{\bi\bk}$ 
in these cases would not be obvious. In addition, this angle might depend on 
the state of the other subsystems, too. Nevertheless, this extended notion 
of the empirical angle may provide the appropriate mathematical formulation 
of Penrose's `ignorance factor' \cite{Pe79,Pe72} between ${\cal S}_{\bi}$ and 
${\cal S}_{\bk}$. 

\item
The operator ${\bf J}_{\bi}\cdot{\bf J}_{\bk}$ was introduced by Moussouris 
in \cite{Mo}. However, in the definition of the empirical angle according to 
him the states had to belong to \emph{finite} dimensional representation 
spaces of $su(2)$. In fact, the denominator in his definition is the 
\emph{norm} of the unbounded operator ${\bf J}_{\bi}\cdot{\bf J}_{\bk}$, which 
is finite only on finite dimensional spaces. In our definition (\ref{eq:2.4}) 
the Hilbert spaces ${\cal H}_{\bi}$ and ${\cal H}_{\bk}$ are \emph{not} 
required to be finite dimensional. Moreover, the geometric idea of angle 
given in the classical theory by (\ref{eq:2.1}) seems to be captured in the 
quantum theory more naturally if, in the denominator, the `lengths' of the 
individual angular momentum vector operators in the given states are used, 
just according to (\ref{eq:2.4}), rather than the norm of ${\bf J}_{\bi}\cdot
{\bf J}_{\bk}$. Indeed, $\cos\theta_{\bi\bi}$ in the state $\vert j_{\bi},j_{\bi}
\rangle$ according to Moussouris would give $(j_{\bi}+1)/j_{\bi}$, which is 
always \emph{greater} than $1$. 

\item
In the theory of canonical quantum gravity, Major \cite{Ma99b} defined 
the angle \emph{operator} acting on two edges of the spin network states, 
labelled by two $su(2)$ Casimir invariants, say $j_{\bi}$ and $j_{\bk}$. 
That operator is ${\bf J}_{\bi}\cdot{\bf J}_{\bk}$ divided by the \emph{norm} 
of ${\bf J}^a_{\bi}$ and of ${\bf J}^a_{\bk}$. Thus, Major's angle operator is 
the correct `operator version' of (\ref{eq:2.1}) (and hence of 
(\ref{eq:2.4})). Nevertheless, since ${\bf J}^a_{\bi}$ and ${\bf J}^a_{\bk}$ 
are \emph{not} bounded, this angle operator is well defined only on 
\emph{finite dimensional} Hilbert spaces. 
\end{enumerate}

In the present paper, we calculate the empirical angle between the subsystems 
only \emph{in pure tensor product states of the composite system} according 
to (\ref{eq:2.4}). Thus, by the first remark above, it is enough to consider 
only two (and, at the end of section \ref{sec-3}, only three) subsystems. In 
the proof 
of the Spin Geometry Theorem, it will be enough to assume that these states 
are tensor products of eigenstates of the Casimir operators of the two 
subsystems, labelled by two Casimir invariants, say $j_1$ and $j_2$. Let 
$\{\vert j_1,m_1\rangle\}$ and $\{\vert j_2,m_2\rangle\}$ be the canonical 
angular momentum bases in the corresponding eigenspaces. Note that here $m$ 
is only an index labeling the vectors of an orthonormal basis in the $2j+1$ 
dimensional carrier space of the unitary representation of $su(2)$, but it 
does \emph{not} refer to any Cartesian frame in the `physical 3-space'. The 
basis $\{\vert j,m\rangle\}$ is chosen to be adapted to the actual choice 
for the components of the vector operator ${\bf J}^a$ \emph{in the abstract 
space of the basic quantum observables}. We choose $\phi_1$ and $\phi_2$ 
simply to be ${\bf U}_1\vert j_1,m_1\rangle$ and ${\bf U}_2\vert j_2,m_2
\rangle$, where the unitary operators ${\bf U}_1$ and ${\bf U}_2$ represent 
$SU(2)$ matrices of the form (\ref{eq:2.2}) with some Euler angles $(\alpha
_1,\beta_1,\gamma_1)$ and $(\alpha_2,\beta_2,\gamma_2)$ in the given 
representations, respectively. 

Then, using ${\bf U}^\dagger{\bf J}^a{\bf U}=R^a{}_b{\bf J}^b$ and how the 
angular momentum operators act on the canonical bases, (\ref{eq:2.4}) yields 
\begin{eqnarray}
\cos\theta_{12}\!\!\!\!&=\!\!\!\!&\frac{\delta_{ab}R^a_1{}_cR^b_2{}_d\langle j_1,
  m_1\vert{\bf J}^c_1\vert j_1,m_1\rangle\langle j_2,m_2\vert{\bf J}^d_2
  \vert j_2,m_2\rangle}{\hbar^2\sqrt{j_1(j_1+1)j_2(j_2+1)}}=\nonumber \\
\!\!\!\!&=\!\!\!\!&\frac{m_1m_2}{\sqrt{j_1(j_1+1)j_2(j_2+1)}}\cos\beta_{12},
  \label{eq:2.5}
\end{eqnarray}
where, by (\ref{eq:2.3}), $\cos\beta_{12}:=(R^{-1}_1R_2)_{33}=\cos
\beta_1\cos\beta_2+\cos(\gamma_1-\gamma_2)\sin\beta_1\sin\beta_2$. 
\medskip

\noindent
Remarks: 
\begin{enumerate}
\item
The expression of $\cos\beta_{12}$ above is a simple consequence of the 
well known addition formulae for the Euler angles, which can be read off 
directly from (\ref{eq:2.3}), too. $\beta_{12}$ is just the angle between 
the unit vectors $(R_1)^a{}_3$ and $(R_2)^a{}_3$, i.e. an angle between 
directions \emph{in the 3-space of the basic quantum observables}. $\theta
_{12}$ depends only on the \emph{relative} orientations of the two subsystems. 

\item
(2.5) shows that, for $\beta_{12}\in[0,\pi/2)$, the empirical angle, $\theta
_{12}$, is \emph{always greater} than $\beta_{12}$, and for $\beta_{12}\in
(\pi/2,\pi]$ it is \emph{always smaller} than $\beta_{12}$. $\theta_{12}=
\beta_{12}$ precisely when $\beta_{12}=\pi/2$. For given $j_1$ and $j_2$, the 
range of $\cos\theta_{12}$ is the \emph{whole} closed interval $[-\sqrt{j_1
j_2/(j_1+1)(j_2+1)},$ $\sqrt{j_1j_2/(j_1+1)(j_2+1)}]$. If $\beta_{12}$ is fixed, 
then the empirical angle is still not fixed and it can take different 
\emph{discrete} values. Note that Planck's constant is canceled from its 
expression. $\theta_{12}$ tends to $\beta_{12}$ \emph{asymptotically} when 
$m_1=j_1$, $m_2=j_2$ and both $j_1$ and $j_2$ tend to infinity. It is this 
limit that is usually considered to be the classical limit of the spin 
systems (see e.g. \cite{Wi}). 

\item
For $\beta_{12}=0$, the empirical angle is given by $\cos\theta^0_{12}=m_1
m_2/\sqrt{j_1(j_1+1)j_2(j_2+1)}$. Here, $m/\sqrt{j(j+1)}$ is just the cosine 
of the `classical' angle between the angular momentum vector of length 
$\sqrt{j(j+1)}$ and its $z$-component with length $m$. Hence, for given 
$m_1m_2\not=0$ and $\beta_{12}=0$, the greater the product $m_1m_2$, the 
smaller the angle $\theta^0_{12}$, but it is \emph{never} zero. Its minimum 
value corresponds to $\cos\theta^0_{12}=\cos\omega_1\cos\omega_2$, where $\cos
\omega:=\sqrt{j/(j+1)}$. Hence $\theta^0_{12}$ is greater than any of $\omega
_1$ and $\omega_2$. $\theta^0_{12}$ tends to zero only asymptotically in the 
$m_1=j_1\to\infty$, $m_2=j_2\to\infty$ (classical) limit. 
\end{enumerate}


\section{The classical limit and the Spin Geometry Theorem}
\label{sec-3}

By (\ref{eq:2.5}), the empirical angles $\theta_{\bi\bk}$ between the 
subsystems ${\cal S}_{\bi}$ and ${\cal S}_{\bk}$ of the composite system in 
the tensor product of the individual states $\vert\phi_{\bi}\rangle=U_{\bi}
\vert j_{\bi},j_{\bi}\rangle$ and $\vert\phi_{\bk}\rangle=U_{\bk}\vert j_{\bk},
j_{\bk}\rangle$, respectively, are given by 
\begin{equation} 
\cos\theta_{\bi\bk}=\sqrt{\frac{j_{\bi}j_{\bk}}{(j_{\bi}+1)(j_{\bk}+1)}}\cos
\beta_{\bi\bk}. \label{eq:3.1}
\end{equation}
Thus, as a consequence of the discussion at the end of the previous section, 
we immediately obtain the following statement: 

\medskip
\begin{proposition} \label{p-3.1}
For arbitrarily small $\epsilon>0$ there is a positive integer $J$ such that, 
in the states above for any $j_{\bi},j_{\bk}>J$, ${\bi},{\bk}=1,\dots,N$, 
$\vert\theta_{\bi\bk}-\beta_{\bi\bk}\vert<\epsilon$ holds. 
\end{proposition}

\noindent
Thus, in the large $j$ limit, the empirical angles $\theta_{\bi\bk}$ tend to
the angles $\beta_{\bi\bk}$ of the three dimensional Euclidean vector space 
of the basic quantum observables. 

However, still we should check that, in this limit, the uncertainties do 
not grow. In fact, we show that these uncertainties tend to zero. First we 
calculate the square of the standard deviation of ${\bf J}_1\cdot{\bf J}_2$. 
The expectation value of its square is 
\begin{eqnarray}
\langle\phi_1\otimes\phi_2\vert({\bf J}_1\cdot{\bf J}_2)^2\vert\phi_1\otimes
  \phi_2\rangle\!\!\!\!&=\!\!\!\!&\langle{\bf J}^a_1\phi_1\vert{\bf J}^c_1
  \phi_1\rangle\delta_{ab}\delta_{cd}\langle{\bf J}^b_2\phi_2\vert{\bf J}^d_2
  \phi_2\rangle= \label{eq:3.2} \\
\!\!\!\!&=\!\!\!\!&(R^{-1}_1R_2)_{ab}(R^{-1}_1R_2)_{cd}\langle j_1,j_1\vert
  {\bf J}^a_1{\bf J}^c_1\vert j_1,j_1\rangle\langle j_2,j_2\vert{\bf J}^b_2
  {\bf J}^d_2\vert j_2,j_2\rangle. \nonumber
\end{eqnarray}
Since the only non-zero matrix elements of ${\bf J}^a{\bf J}^b$ in the states 
$\vert j,j\rangle$ are 
\begin{eqnarray*}
&{}&\langle j,j\vert{\bf J}^1{\bf J}^1\vert j,j\rangle=\langle j,j\vert
 {\bf J}^2{\bf J}^2\vert j,j\rangle=\frac{1}{2}\hbar^2j, \\
&{}&\langle j,j\vert{\bf J}^1{\bf J}^2\vert j,j\rangle=-\langle j,j\vert
 {\bf J}^2{\bf J}^1\vert j,j\rangle=\frac{\rm i}{2}\hbar^2j, \\
&{}&\langle j,j\vert{\bf J}^3{\bf J}^3\vert j,j\rangle=\hbar^2j^2,
\end{eqnarray*}
(\ref{eq:3.2}) takes the form 
\begin{eqnarray*}
\langle\phi_1\otimes\phi_2\vert\!\!\!\!&{}\!\!\!\!&({\bf J}_1\cdot{\bf J}_2)^2
  \vert\phi_1\otimes\phi_2\rangle=\frac{1}{4}\hbar^4j_1j_2\Bigl(\bigl((R^{-1}_1
  R_2)_{11}\bigr)^2+\bigl((R^{-1}_1R_2)_{12}\bigr)^2+\bigl((R^{-1}_1R_2)_{21}
  \bigr)^2+ \\
+\!\!\!\!&{}\!\!\!\!&\bigl((R^{-1}_1R_2)_{22}\bigr)^2\Bigr)+\frac{1}{2}\hbar^4
  (j_1)^2j_2\Bigl(\bigl((R^{-1}_1R_2)_{31}\bigr)^2+\bigl((R^{-1}_1R_2)_{32}\bigr)
  ^2\Bigr)+ \\
+\!\!\!\!&{}\!\!\!\!&\frac{1}{2}\hbar^4j_1(j_2)^2\Bigl(\bigl((R^{-1}_1R_2)_{13}
  \bigr)^2+\bigl((R^{-1}_1R_2)_{23}\bigr)^2\Bigr)+\hbar^4(j_1)^2(j_2)^2\bigl(
  (R^{-1}_1R_2)_{33}\bigr)^2+ \\
+\!\!\!\!&{}\!\!\!\!&\frac{1}{2}\hbar^4j_1j_2\Bigl((R^{-1}_1R_2)_{12}(R^{-1}_1R
  _2)_{21}-(R^{-1}_1R_2)_{11}(R^{-1}_1R_2)_{22}\Bigr).
\end{eqnarray*}
Using the explicit form (\ref{eq:2.3}) of the rotation matrix, a lengthy but 
elementary calculation yields that 
\begin{eqnarray*}
\langle\phi_1\otimes\phi_2\vert\!\!\!\!&{}\!\!\!\!&({\bf J}_1\cdot{\bf J}_2)^2
  \vert\phi_1\otimes\phi_2\rangle=\frac{1}{4}\hbar^4j_1j_2\bigl(1+\cos^2\beta
  _{12}\bigr)+\frac{1}{2}\hbar^4(j_1)^2j_2\bigl(1-\cos^2\beta_{12}\bigr)+ \\
+\!\!\!\!&{}\!\!\!\!&\frac{1}{2}\hbar^4j_1(j_2)^2\bigl(1-\cos^2\beta_{12}\bigr)
  +\hbar^4(j_1j_2)^2\cos^2\beta_{12}-\frac{1}{2}\hbar^4j_1j_2\cos\beta_{12}.
\end{eqnarray*}
Hence, the square of the standard deviation of ${\bf J}_1\cdot{\bf J}_2$ in 
the state $\phi_1\otimes\phi_2$ is 
\begin{eqnarray*}
\bigl(\Delta_\phi({\bf J}_1\cdot{\bf J}_2)\bigr)^2\!\!\!\!&=\!\!\!\!&
  \langle\phi\vert({\bf J}_1\cdot{\bf J}_2)^2\vert\phi\rangle-\Bigl(
  \langle\phi\vert{\bf J}_1\cdot{\bf J}_2\vert\phi\rangle\Bigr)^2= 
  \nonumber \\
\!\!\!\!&=\!\!\!\!&\frac{1}{2}\hbar^4j_1j_2\Bigl(\frac{1}{2}\bigl(1-\cos
  \beta_{12}\bigr)^2+(j_1+j_2)\sin^2\beta_{12}\Bigr). 
\end{eqnarray*}
Since the state $\phi_1\otimes\phi_2$ is an eigenstate both of $({\bf J}_1)
^2$ and $({\bf J}_2)^2$, finally we obtain that the square of the uncertainty 
of $\cos\theta_{\bi\bk}$ in the state $\phi=\phi_1\otimes\cdots\otimes\phi_N$, 
defined by the first equality below, is 
\begin{equation}
\bigl(\Delta_\phi\cos\theta_{\bi\bk}\bigr)^2:=
\frac{\bigl(\Delta_\phi({\bf J}_{\bi}\cdot{\bf J}_{\bk})\bigr)^2}{\langle\phi
\vert({\bf J}_{\bi})^2\vert\phi\rangle\,\langle\phi\vert({\bf J}_{\bk})^2\vert
\phi\rangle}=
\frac{1}{4}\frac{(1-\cos\beta_{\bi\bk})^2+2(j_{\bi}+j_{\bk})\sin^2\beta_{\bi\bk}}
{(j_{\bi}+1)(j_{\bk}+1)}. \label{eq:3.4}
\end{equation}
For given $j_{\bi}$ and $j_{\bk}$ this uncertainty is zero precisely at $\beta
_{\bi\bk}=0$, and it takes its maximal value at $\cos\beta_{\bi\bk}=1/(1-2(j
_{\bi}+j_{\bk}))$. However, independently of $\beta_{\bi\bk}$, this uncertainty 
tends to zero if $j_{\bi},j_{\bk}\to\infty$. With this conclusion we have 
proven the next statement. 

\medskip
\begin{proposition} \label{p-3.2}
For arbitrarily small $\epsilon>0$ there is a positive integer $J$ such that, 
in the states above for any $j_{\bi},j_{\bk}>J$, ${\bi},{\bk}=1,\dots,N$, 
$\Delta_\phi\cos\theta_{\bi\bk} <\epsilon$ hold. 
\end{proposition}

\medskip
\noindent
With the propositions \ref{p-3.1} and \ref{p-3.2} at hand we have already 
given a simple proof of (a version of) the Spin Geometry Theorem in the 
traditional framework of quantum mechanics: 

\begin{theorem*}
Let ${\cal S}$ be composed of the quantum mechanical systems ${\cal S}_1,
...,{\cal S}_N$. Then there is a collection of pure tensor product states 
of ${\cal S}$, $\phi_1(j_1)\otimes\cdots\otimes\phi_N(j_N)$ indexed by an 
$N$-tuple $(j_1,...,j_N)$ of non-negative integers or half-odd-integers, 
such that, in the $j_1,...,j_N\to\infty$ limit, the empirical angles between 
these subsystems converge with asymptotically vanishing uncertainty to angles 
between directions of the three dimensional Euclidean vector space. 
\end{theorem*}

\noindent
Remarks: 
\begin{enumerate}
\item
Since for any given ${\bi}$ and $j'_{\bi}\not=j_{\bi}$ the states $\vert j
_{\bi},j_{\bi}\rangle$ and $\vert j'_{\bi},j'_{\bi}\rangle$ belong to orthogonal 
subspaces of the Hilbert space ${\cal H}_{\bi}$ of all the pure states of the 
system ${\cal S}_{\bi}$, the states ${\bf U}_1\vert j_1,j_1\rangle\otimes
\cdots\otimes{\bf U}_N\vert j_N,j_N\rangle\in{\cal H}_1\otimes\cdots\otimes
{\cal H}_N$ labelled by different $N$-tuples, say $(j_1,j_2,...,j_N)$ and 
$(j'_1,j_2,...,j_N)$, are orthogonal to one another. Hence, the sequence of 
the states $\phi_1(j_1)\otimes\cdots\otimes\phi_N(j_N)\in{\cal H}_1\otimes
\cdots\otimes{\cal H}_N$ does \emph{not} converge to any normalized state in 
the strong topology of ${\cal H}_1\otimes\cdots\otimes{\cal H}_N$. (In the 
weak topology, it converges to zero.) Therefore, there is \emph{no} quantum 
state of the system which would represent the above classical limit $j_1,...,
j_N\to\infty$. It is not clear whether or not one can find \emph{actual 
states} in ${\cal H}_1\otimes\cdots\otimes{\cal H}_N$, analogous e.g. to the 
so-called canonical coherent states of the Heisenberg systems and which 
could also be interpreted as the composite system's `most classical state' 
(see e.g. \cite{Sz21a}), in which the empirical angles would coincide with 
those of the three dimensional Euclidean vector space. 

\item 
One novelty of the analysis behind the above version of the Spin Geometry 
Theorem is that it is based on a concept of empirical angle, viz. that given 
by (\ref{eq:2.4}), which is well defined not only asymptotically (like that 
in the version of the Spin Geometry Theorem proven in \cite{Mo}), but 
\emph{even at the genuine quantum level}. Thus, in the present approach, 
some non-trivial aspect of the \emph{quantum geometry} defined by the 
quantum mechanical systems is already shown up. (We discuss this issue a bit 
more in the next section.) The other novelty is that it gives 
\emph{explicitly} a sequence of states which provides the correct, expected 
classical limit. 

It might be worth noting that \emph{mathematically} the Theorem stated in 
\cite{Pe79,Pe71a,Pe10,Pe71b,Pe72}, the version proven in \cite{Mo} and the 
version above are \emph{not} equivalent. Nevertheless, their \emph{physical 
content}, viz. that the conformal structure of the Euclidean 3-space can be
recovered in the classical limit from quantum observables, is the same. 
Hence we can consider them only different \emph{versions} of the \emph{same} 
physical theorem. 

\item
According to expectations of certain recent investigations (see e.g. 
\cite{Ra,Ja,CaCaMi}), the geometry of 3-space/spacetime emerges from the 
\emph{entanglement} of the \emph{states} of the quantum subsystems of the 
Universe. However, the results of the present work do not seem to support 
this expectation, at least in the quantum \emph{mechanical} approximation 
of the quantum world. In fact, the Spin Geometry Theorem could successfully 
be proven using only \emph{pure tensor product states} of the subsystems. 
The entanglement of the \emph{states} of the subsystems was not needed. On 
the other hand, the quantum operator ${\bf J}_{\bi}\cdot{\bf J}_{\bk}$, by 
means of which the empirical angles were defined, is \emph{a sum of the 
products} of observables, ${\bf J}^1_{\bi}{\bf J}^1_{\bk}+{\bf J}^2_{\bi}
{\bf J}^2_{\bk}+{\bf J}^3_{\bi}{\bf J}^3_{\bk}$, which structure is analogous 
to that of the entangled states. But, in contrast to local quantum 
\emph{field theory}, in quantum \emph{mechanics} there is no locality: the 
quantum operators `feel' the whole wave function on the \emph{entire} 
momentum/configuration space. Therefore, the entanglement can be considered 
to be built already into the structure of the \emph{quantum mechanical 
observables of the composite system}, by means of which the geometry of 
3-space/spacetime can be defined in an operational way. The \emph{states} 
do not need to be entangled. 

\item
Using the natural volume 3-form $\varepsilon_{abc}$ on the algebra $su(2)$ 
of the basic quantum observables, a further potentially interesting 
geometric notion, viz. the `empirical 3-volume elements' can be introduced. 
In the state $\phi\in{\cal H}$ spanned by three subsystems this is defined 
by 
\begin{equation}
v_{\bi\bj\bk}:=\frac{1}{3!}\varepsilon_{abc}\frac{\langle\phi\vert{\bf J}^a_{\bi}
    {\bf J}^b_{\bj}{\bf J}^c_{\bk}\vert
\phi\rangle}{\sqrt{\langle\phi({\bf J}_{\bi})^2\vert\phi\rangle}\sqrt{\langle
\phi({\bf J}_{\bj})^2\vert\phi\rangle}\sqrt{\langle\phi({\bf J}_{\bk})^2\vert
\phi\rangle}}. \label{eq:3.5}
\end{equation}
Then, in the tensor product of the states of the form ${\bf U}\vert j,m
\rangle$ with the unitary operator ${\bf U}$ representing some $SU(2)$ 
matrix $U^A{}_B$ of the form (\ref{eq:2.2}), this expression gives 
\begin{equation}
v_{\bi\bj\bk}=\frac{m_{\bi}m_{\bj}m_{\bk}}{\sqrt{j_{\bi}(j_{\bi}+1)j_{\bj}(j_{\bj}+1)
j_{\bk}(j_{\bk}+1)}}\frac{1}{3!}\varepsilon_{abc}(R_{\bi})^a{}_3(R_{\bj})^b{}_3
(R_{\bk})^c{}_3, \label{eq:3.6}
\end{equation}
where $R^a{}_b$ is the rotation matrix (\ref{eq:2.3}) corresponding to 
$U^A{}_B$, and the second factor in (\ref{eq:3.6}) is just the Euclidean 
3-volume of the tetrahedron spanned by the unit vectors $(R_{\bi})^a{}_3$, 
$(R_{\bj})^a{}_3$ and $(R_{\bk})^a{}_3$. Thus, even if $m_{\bi}=j_{\bi}$, $m_{\bj}
=j_{\bj}$ and $m_{\bk}=j_{\bk}$, $v_{\bi\bj\bk}$ is \emph{always smaller} than its 
Euclidean counterpart: the former is only \emph{conformal} to the latter, 
and it tends to the Euclidean 3-volume only in the $m_{\bi}=j_{\bi},m_{\bj}=
j_{\bj},m_{\bk}=j_{\bk}\to\infty$ limit. 

\end{enumerate}


\section{A classical model of the `space of the quantum directions'}
\label{sec-4}

In \cite{Pe72}, Penrose (quoting Aharonov, too) raises the possibility in 
the context of the quantum mechanical double slit experiment that the `true' 
geometry that the electron `sees' might be different from the `classical' 
geometry of the two slits. Motivated by this idea, we may ask `What kind of 
geometry should we have if we want to arrange the ``empirical'' geometric 
notions and quantities introduced via the observables of the quantum 
systems?' In particular, what could be the geometry in which the empirical 
angles and 3-volume elements, defined by the composite quantum system 
${\cal S}={\cal S}_1\cup\cdots\cup{\cal S}_N$, are angles and 3-volume 
elements? 

To illustrate by a classical model how this `true' (conformal) geometry 
might look like, for the sake of simplicity suppose that all the subsystems 
have the same total angular momentum $j$. Let us assume that the empirical 
angles are angles between pairs of unit vectors of (and the empirical 
3-volumes are 3-volumes of tetrahedra formed by triplets of unit vectors 
in) some $n$ dimensional real vector space. This vector space is modeled by 
$\mathbb{R}^n$, which is endowed by some positive definite metric $G
_{\alpha\beta}$, $\alpha,\beta=1,...,n$. 

By (\ref{eq:2.5}) the empirical angle between the `directions' of two 
subsystems \emph{cannot be smaller} than $\theta^{min}=\arccos(j/(j+1))$
and \emph{cannot be greater} than $\theta^{max}=\arccos(-j/(j+1))=\pi-\theta
^{min}$. (For example, for $j=1/2$ these bounds are $\approx70.53^\circ$ and 
$\approx109.47^\circ$; and for $j=1$ these are $60{}^\circ$ and $120{}^\circ$, 
respectively.) Clearly, for any given $N$, these empirical angles can always 
be arranged in $\mathbb{R}^n$ for large enough but finite $n$. Nevertheless, 
the existence of the bounds $\theta^{min}$ and $\theta^{max}$ provides a lower 
bound for $n$. (For a given state $\phi_1\otimes\cdots\otimes\phi_N$, the 
optimal value of $n$ might be determined by a procedure analogous to that in 
the so-called sphere packing problem \cite{ConSlo}, see below.) Let the 
unit vector $V^\alpha _{\bi}$ represent the `direction' associated with the 
${\bi}$th subsystem in this space. Then let us draw two solid cones in 
$\mathbb{R}^n$ with $V^\alpha_{\bi}$ as their common axis, their vertices at 
the origin, and with the opening angle $\theta^{min}$ and $\theta^{max}$, 
respectively, such that the cone with opening angle $\theta^{max}$ contains 
the cone with opening angle $\theta^{min}$. Then the result that the empirical 
angle between the `directions' of any two subsystems cannot be greater than 
$\theta^{max}$ and cannot be smaller than $\theta^{min}$ implies that the 
\emph{inner cone} with axis $V^\alpha_{\bi}$ and that with axis $V^\alpha_{\bk}$ 
intersect each other \emph{at most} along one line in their lateral surface, 
and their \emph{outer cone} intersect each other \emph{at least} along one 
line in their lateral surface. 

Let $p_{\bi}$ be the point of the unit sphere $S^{n-1}$ in $\mathbb{R}^n$ that 
the unit vector $V^\alpha_{\bi}$ defines, let $C_{\bi}$ denote the intersection 
of the cone with axis $V_{\bi}$ and opening angle $\theta^{max}$ with the unit 
sphere, and let $c_{\bi}$ be the intersection of the inner cone with the unit 
sphere. Thus, $C_{\bi}$ and $c_{\bi}$ are concentric spherical caps with the 
common centre $p_{\bi}$ on $S^{n-1}$. Then the resulting spherical caps 
$C_{\bi}$ and $C_{\bk}$, ${\bi}\not={\bk}$, \emph{must} intersect each other 
at least in one point, but the corresponding inner spherical caps $c_{\bi}$ 
and $c_{\bk}$ \emph{may} intersect each other at most in one point. Therefore, 
the set of the empirical angles between any two subsystems considered in 
section \ref{sec-2} may be represented by the set of these configurations of 
the points $p_{\bi}$ and the corresponding pairs $(C_{\bi},c_{\bi})$, ${\bi}=1,
...,N$. There are \emph{no distinguished directions} in this space, i.e. 
the space does \emph{not} have any naive lattice structure, \emph{but the 
angle between any two directions cannot be arbitrarily small or arbitrarily 
large}. 

This second realization of the classical model of the `space of the quantum 
directions' makes it possible, at least in principle, to determine the 
minimal dimension $n$ in which the directions of $N$ subsystems, each with 
spin $j$, can be arranged: this is the minimal dimension for which $N$ pairs 
of concentric $(n-1)$ dimensional spherical caps, or rather balls, with 
given radii can be packed into the unit sphere $S^{n-1}$ satisfying the above 
conditions. This is a version of the `sphere packing problem' of 
\cite{ConSlo}. 

Requiring that the 3-volume elements determined by $V^\alpha_{\bi},V^\alpha
_{\bj}$ and $V^\alpha_{\bk}$ in $\mathbb{R}^n$ be just $v_{\bi\bj\bk}$ for the 
composite system, (\ref{eq:3.6}) might suggest to choose the metric $G
_{\alpha\beta}$ to be conformal to the Euclidean one, $G_{\alpha\beta}=(j/(j+1))
\delta_{\alpha\beta}$.

\section{Acknowledgments}
\label{sec-5}

Thanks are due to Paul Tod for making the dissertation \cite{Mo} available 
to me (prior the appearance of the link to it); and to P\'eter Vecserny\'es 
for the numerous discussions on the algebraic formulation of quantum theory 
and for calling my attention to the sphere packing problem \cite{ConSlo}. I 
am grateful to both of them for the careful reading of an earlier version 
of the present paper and for their helpful remarks and suggestions for its 
improvement. Thanks are also due to Ted Jacobson for the discussion on the 
role of entanglement in the emergence of the geometry of the 
3-space/spacetime from quantum mechanics, and also for the link in reference 
\cite{Mo}; to Roger Penrose for his remarks on the history of his original
theorem as well as for the reference \cite{Pe10}; and to J\"org Frauendiener 
for making \cite{Pe10} available to me. 

No funds, grants or support was received.

\end{document}